\documentclass[fleqn,twoside]{article}

\newif\ifpreprint
\preprinttrue

\usepackage{espcrc2}
\usepackage[dvips]{graphicx}
\ifpreprint
\title{
\rightline{\normalsize FERMILAB-Conf-02/236-E}
\vspace{0.2cm}
First Observation of a Family of Double-Charm 
Baryons\thanks{Talk at the 
31st International Conference on High Energy Physics (ICHEP 2002),
Amsterdam, July 24 - 31, 2002. To be published in the proceeding.}}
\else
\title{First Observation of a Family of Double-Charm Baryons}
\fi
\author{J. S. Russ\address{Carnegie Mellon University
 \\ Pittsburgh, PA 15213 USA} \\ on behalf of the 
{SELEX Collaboration}\thanks{SELEX Collaboration:
Ball State University,
Bogazici University,
Carnegie Mellon University,
Centro Brasileiro de Pesquisas Fisicas (Rio de Janeiro),
Fermi National Accelerator Laboratory,
Petersburg Nuclear Physics Institute,
IHEP (Beijing),
IHEP (Protvino),
ITEP (Moscow),
Moscow State University,
Max-Planck-Institut f\"ur Kernphysik (Heidelberg),
Tel Aviv University,
Universidad Aut\'onoma de San Luis Potos\'{\i},
University of Bristol,
Universidade Federal da Paraiba,
University of Iowa,
University of Michigan--Flint,
University of Rome ``La Sapienza'', INFN,
University of S\~ao Paulo,
University of Trieste.}
}

\begin{document}

\begin{abstract}
   The SELEX experiment (E781) at Fermilab has candidates for high
mass states decaying to $\Lambda_c^+ \rm{K}^- \pi^+$ and $\Lambda_c^+ \rm{K}^- \
pi^+ \pi^+$, Cabibbo-allowed decay modes of doubly-charmed baryons $\Xi_{cc}^+$ and
$\Xi_{cc}^{++}$.  The masses are consistent with
theoretical considerations, but the spectroscopy is surprising.
Limited lifetime information suggests that 
$\tau_{\Xi_{cc}^{++}} \sim \tau_{\Xi_{cc}^{+}}$.
\vspace{1pc}
\end{abstract}

\maketitle

\section{Introduction}

The existence of baryons with two and three charm quarks is expected from our
present understanding of hadronic structure.  In the double-charm system one
expects a J=1/2 ground state iso-doublet, termed $\Xi_{cc}^{+,++}$ in PDG notation~\cite{PDG}.  Most predictions for the masses of the J=1/2 states and the
J=3/2 hyperfine excitations expect the ground state near 3.6 GeV/$\rm{c}^2$
 and a hyperfine split of 60 MeV/$\rm{c}^2$~\cite{bphys}.

The production mechanism for double-charm states is not clear.  Perturbative pictures
treat production as successive c$\overline{\rm{c}}$ pair production with enough spatial overlap
to form hadrons.  Such cross sections are small compared to single-charm production.
  
\section{Features of the Selex spectrometer}

The SELEX experiment at Fermilab is a 3-stage magnetic spectrometer
~\cite{spec}.  The negative 600 GeV/c Fermilab Hyperon Beam
had about equal fluxes of $\pi^-$ and $\Sigma^-$.  The positive beam was
92\% protons. For 
charm momenta in a range of 100-500 GeV/c mass resolution is constant
and primary (secondary) vertex resolution is typically 270 (560) $\mu$m.
A RICH detector labelled all particles above 25 GeV/c~\cite{RICH}.  Details 
of single-charm
analyses involving $\Lambda _{c} ^{+}  \rightarrow p K ^{-} \pi ^{+}$
reconstructions can be found in ~\cite{lclife,lcprod}.  The double-charm
search discussed here began with the sample of 1630 $\Lambda_c^+$ events
used in the lifetime analysis~\cite{lclife}.

\section{Double-charm Analysis}

\begin{figure}[h]
%\epsfxsize120pt
%\figurebox{}{}{ccdfig.eps}
%\psfig{figure=ccdfig.eps,width=80mm}
\includegraphics[width=7cm]{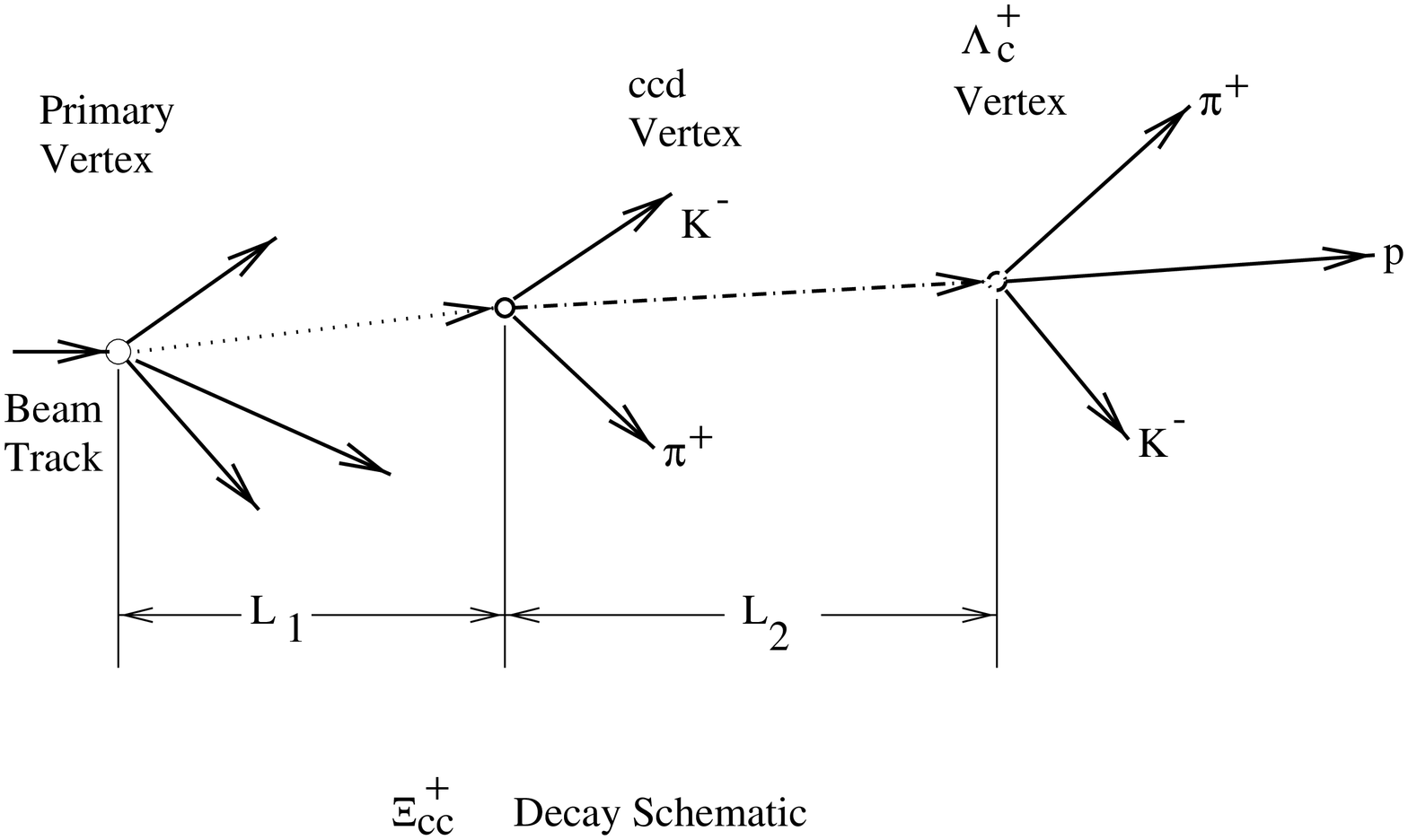}
\caption{Schematic of $\Xi_{cc}^{+} \rightarrow \rm{K}^- \pi^+  \Lambda_c^+$ } \label{fig:ccdfig}
\end{figure}

This topological search for double-charm baryons asks for a decay vertex lying between
the primary vertex and the observed $\Lambda_c^+$ decay vertex.    A Cabibbo-allowed
$\Xi_{cc}^+$ decay can give a final-state $\Lambda_c^+$, a $\rm{K}^-$, and a $\pi^+$, 
shown schematically in Fig.~\ref{fig:ccdfig}.

We reconstruct total charge states Q=1 (neutral 2-prong vertex) and Q=2
(positive 3-prong vertex) in separate reconstructions.  Few of the tracks from
 the intermediate vertex  are RICH-
identified.  We call the negative track a kaon in the right-sign
reconstruction. The same events have a  wrong-sign
reconstruction, calling the negative track a pion.

Event selection cuts used here were taken without change from previous 
single-charm studies.  For short-lived states, $\rm{L_1}/\sigma_1 \ge 1$ and the $\Lambda_c^+$
momentum vector must point back to the primary vertex within a $\chi^2$ cut of 4.
We have varied the cuts and observe that no signal significance depends
critically on any cut value.  The cuts were checked with simulation 
studies using several production models for double charm.

\subsection{Q = 1 reconstruction}

\begin{figure}[h]
\includegraphics*[width=70mm]{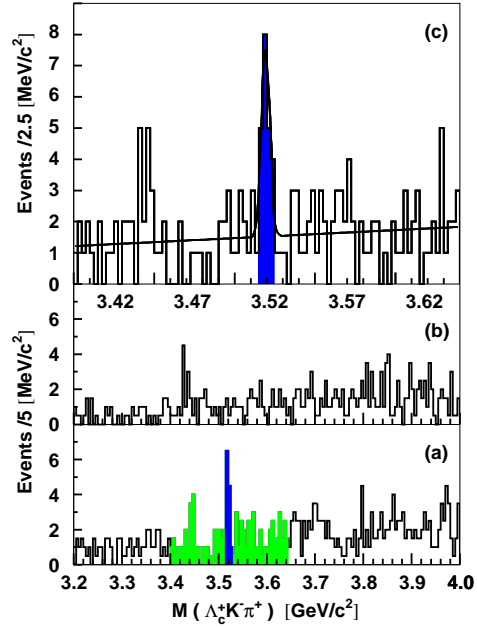}
\caption{(a) The $\Lambda_c^+ K^- \pi^+$ mass distribution in 5 MeV/$c^2$ bins.
             The shaded region is shown in more detail in (c).
         (b) The wrong-sign combination  $\Lambda_c^+ K^+ \pi^-$ mass distribution in 5 
MeV/$c^2$ bins.
         (c) The signal (shaded) region (22 events) and sideband mass 
             regions (140 events) in 2.5 MeV/$\rm{c}^2$ bins.  
             The fit is a Gaussian plus linear background.}
\label{fig2}
\end{figure}

For single-charged baryons, the $ \rm{K}^- \pi^+  \Lambda_c^+$ mass distribution is
shown in Fig.~\ref{fig2}.
Fig.~\ref{fig2}(c) shows a $\Xi_{cc}^{+}$ candidate at 3520 MeV/$\rm{c}^2$,
consistent with most model calculations.  The peak is narrow but consistent with simulation.
The general agreement between right-sign (a,c) and wrong-sign (b) average levels and fluctuations
in Fig.~\ref{fig2} confirms that most events are combinatoric background.  The
signal region in (c) contains 22 events with a background of 6.1 $\pm$ 0.51 events, for a single-bin 
significance of 6.3 $\sigma$.   The probability of such an 
excess is less than $10^{-6}$ for a single bin.   We searched for a peak in the interval 
3.2-4.3 GeV/$\rm{c}^2$, or 110 bins.  The probability of such a fluctuation anywhere in the search
interval is $ < 1.1 \times 10^{-4}$.  SELEX has reported this as the first
observation of a doubly-charmed baryon~\cite{prl}.

Using the reduced proper decay time  
$ t^{*} ={M(L-L_{min})/ p c }$ we find that the lifetime $ < $ 33 fs at 90\%
confidence.  Here $\rm{L_{min}} = \sigma_1$ the error on the vertex separation
$\rm{L_1}$.   The SELEX proper time resolution is about 20 fs.  With 
constructive interference between
the two c-quark decay amplitudes along with the W-exchange amplitudes, this 
state could have a lifetime shorter than the $\Xi_c^0$.

\subsection{Q=2 Reconstruction}

\begin{figure}[htb]
%\epsfxsize120pt
\includegraphics*[width=70mm]{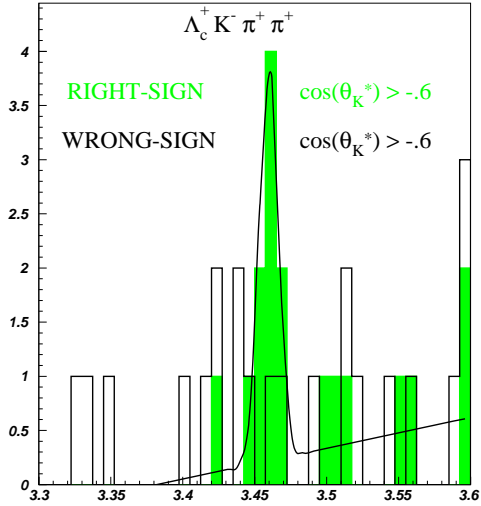}
\caption{ $\Xi_{cc}^{+} \rightarrow \rm{K}^- \pi^+ \pi^+ \Lambda_c^+$  mass distribution
in 7.5 MeV/$\rm{c}^2$ bins.  Signal events are shaded.  Wrong-sign background is shown as open histogram
boxes.   } \label{m6lo}
\end{figure}

For double-charged baryons, we look for an isospin partner of
the $\Xi_{cc}^+$(3520).  The $ \rm{K}^- \pi^+ \pi^+  \Lambda_c^+$ mass
distribution  in the vicinity of 3520 MeV/$\rm{c}^2$ is
shown in Fig.~\ref{m6lo}, along with the wrong-sign background.

There is a $\Xi_{cc}^{++}$ candidate at 3460 MeV/$\rm{c}^2$.
The observed width (6 $\pm$ 1 MeV/$\rm{c}^2$) matches simulation.
Events outside the signal region show a strong preference for the
center-of-mass (CM) angle of the negative track to be near 180 degrees.
Simulation indicates that a cut to remove such events should have
very little effect on the signal region for a phase-space decay distribution.
That is indeed the case in the data.  With the selection shown, 
we find 9 events in the peak, compared to an
expected background of 1 event.  The Poisson probability that there is
an excess of 8 events or more anywhere on the plot is $10^{-5}$.
There are too few events to attempt a lifetime analysis.  The 
$\Xi_{cc}^{++}$ candidates have a raw average proper time (uncorrected
for acceptance) comparable
to that for the $\Xi_{cc}^+$ candidates, suggesting that 
$\tau_{\Xi_{cc}^{++}} \sim \tau_{\Xi_{cc}^{+}}$.
 
\subsection{Production }

Both $\Xi_{cc}^+$ and $\Xi_{cc}^{++}$ states are produced only by baryon beams in 
SELEX data.  There are no signal candidates from the  pion beam.   Simulation studies 
suggest that the double-charm states may account for as
much as 40\% of the $\Lambda_c^+$ sample seen in this experiment, a surprisingly high
fraction.     This situation is reminiscent of the discovery of the $\Xi_c^+$
baryon in the WA62 experiment at CERN, using a 135 GeV hyperon beam~\cite{biagi}.  
The FOCUS photoproduction experiment at Fermilab has looked for these 
states using their $\Lambda_c^+$ events and sees no signal 
peaks~\cite{FOCUS}.  If SELEX is correct, the hadroproduction mechanism
is unusual.

\section{Summary}

SELEX has introduced two statistically-compelling new 
high-mass states that decay into a final state $\Lambda_c^+$, $\rm{K}^-$ and one or two 
$\pi^+$, as expected for double-charm baryon decays.  The 3520
MeV/$\rm{c}^2$ state satisifies all expectations for being a $\Xi_{cc}^{+}$
state~\cite{prl}.  Its lifetime is shorter than 30 fs
at 90\% confidence.  The 3460 MeV/$\rm{c}^2$ state has the decay characteristics of
a $\Xi_{cc}^{++}$ state.  It is difficult to understand the 60  MeV/$\rm{c}^2$
mass difference between the Q=1 and Q=2 states if they are members of the ground
state isodoublet.  However, any other interpretation is also problematic.

\end{document}